\documentstyle[12pt]{article}
\newcommand{\be}[1]{
\begin{eqnarray}\label{#1}}
\newcommand{\ee}{\end{eqnarray}}
\newcommand{\ci}[1]{\cite{#1}}
\newcommand{\re}[1]{(\ref{#1})}
\newcommand\pbar{\bar{\psi}}
\newcommand\p{\psi }

\def\xslash{\rlap/{\mkern-1mu x}}
\def\zslash{\rlap/{\mkern-1mu z}}
\def\pslash{\rlap/{\mkern-1mu p}}
\newcommand{\pert}{{\cal D}}

\newcommand{\spin}[1]{ \langle\mskip-3mu \langle{#1}
\rangle\mskip-3mu\rangle}
\begin{document}
\renewcommand{\thefootnote}{\fnsymbol{footnote}}
\begin{flushright}
\begin{tabular}{l}
RUB-TP2-19/00\\ TPR-00-15\\
\end{tabular}
\end{flushright}
\begin{center}
{\bf\Large DVCS on the nucleon to the twist-3 accuracy}

\vspace{0.5cm} N. Kivel$^{a,b}$,  M.V. Polyakov$^{b,c}$

\begin{center}
{\em $^a$ Institut f\"ur Theoretische Physik, Universit\"at
Regensburg \\ D-93040 Regensburg, Germany \\ $^b$Petersburg
Nuclear Physics Institute, 188350, Gatchina, Russia\\ $^c$
Institut f\"ur Theoretische Physik II, Ruhr-Universit\"at
Bochum,\\ D-44780 Bochum, Germany }

\end{center}
\end{center}

\begin{abstract}
The amplitude of the deeply virtual Compton scattering off nucleon
is computed to the twist-3 accuracy in the Wandzura-Wilczek (WW)
approximation. The result is presented in the form which can be
easily used for analysis of DVCS observables.
\end{abstract}

\section*{\normalsize \bf Introduction}

Deeply virtual Compton scattering (DVCS) \cite{DVCS1,DVCS2} is the
cleanest hard process which is sensitive to the skewed parton
distributions (SPD) and has been the subject of extensive
theoretical investigations for a few years. First experimental
data became recently available (see e.g. \cite{exp1,exp2,amarian})
and much more data are expected from JLAB, DESY, and CERN in the
near future. The leading order ($1/Q^0$) DVCS amplitude owing to
factorisation theorems \cite{Rad97,Ji98,Col99} can be expressed in
terms of twist-2 skewed parton distributions. However, as the
typical experimentally accessible $Q^2$ are by no means large,
 studies of the power (higher twist) corrections to the DVCS amplitude
are very important. The leading power corrections to the DVCS
amplitude are of the order $1/Q$ only (hence  twist-3) and their
effect on the DVCS observables can be rather sizeable. Also the
size of the twist-3 corrections increases with the increase of the
momentum transfer squared $t$ because they enter the amplitude
typically as $\sqrt{-t}/Q$. Therefore, taking into account  the
twist-3 corrections is important for the problem of continuation
of the twist-2 SPD's to $t=0$.

Recently the DVCS amplitude  was computed in
Refs.~\cite{Anikin,Penttinen,BM,RW1} including the terms of the
order $O(1/Q)$. The inclusion of such terms is mandatory to ensure
the electromagnetic gauge invariance of the DVCS amplitude to the
order $\sqrt{-t}/Q$. At the order $1/Q$ the DVCS amplitude depends on a
set of new skewed parton distributions.
 Recently it was shown in
Ref.~\cite{BM} that in the so-called Wandzura-Wilczek (WW)
approximation these new functions can be expressed in terms of
twist-2 SPD's.

The WW approximation corresponds to neglecting  the nucleon
matrix elements of the `genuine twist-3' operators of the type
$\bar \psi G \psi$, {\it i.e.} neglecting the nonperturbative
quark-gluon correlations in the nucleon. In principle this
approximation is not justified by any small parameter of the
theory. However, recent measurements of the polarized structure
function $g_T(x)$ \cite{g2izmer} show that the WW approximation
works pretty well in this case. Also the estimates of the `genuine
twist-3' contributions to polarized structure functions $g_T(x)$
and $h_L(x)$ in the theory of instanton vacuum \cite{BPW,DP}
(which is a model of non-perturbative quark-gluon correlations)
showed that these contributions are parametrically suppressed
relative to the `kinematical' part of the twist-3 contributions by
the packing fraction of instantons in the vacuum.

Generically the new twist-3 SPD's in the WW approximation posses
discontinuities  at the points $x=\pm\xi$ \cite{RW1,KPST} which
potentially, may lead to divergencies in the DVCS amplitude and
hence to the violation of the factorisation at the $1/Q$ accuracy.
However, it was shown in \cite{RW1,KPST} for the case of DVCS on
the pion target that these dangerous divergencies are cancelled to
the $1/Q$ accuracy. In Ref.~\cite{KPST} general arguments were
given that the cancellation of the divergencies should persist
also for any target. In the present paper we demonstrate by
explicit calculations that indeed the DVCS amplitude on the
nucleon to the $1/Q$ accuracy in the WW approximation is free of
the divergencies. The results can easily be used for the analysis
of the DVCS observables. Collection of formulae the reader can
find in the Appendix.

\section*{\normalsize \bf DVCS amplitude on the nucleon at the twist-3
accuracy}

We shall work in the reference frame where the average nucleon
momenta and the virtual photon momentum are collinear and have
opposite directions. For convenience we introduce light-cone
vectors $n,\, n^*$ such that \be{nnstar} n\cdot n=0, \, n^*\cdot
n^*=0, n\cdot n^*=1. \ee We also define the  transverse metric and
antisymmetric transverse epsilon tensor \footnote{The Levi-Civita
tensor $\epsilon_{\mu \nu \alpha\beta}$ is defined as the totally
antisymmetric tensor with $\epsilon_{0123}=1$ }:
\be{gt} (-g^{\mu
\nu})_\perp = -g^{\mu \nu}+ n^\mu n^{*\, \nu}+n^\nu n^{*\, \mu},
\quad \epsilon^\perp_{\mu \nu}= \epsilon_{\mu \nu
\alpha\beta}n^\alpha n^{*\,\beta}\, . \ee For an arbitrary vector
$a_\mu$ we will use the shorthand notation \be{defdot}
 a_+\equiv a_\mu n^\mu, \quad  a_-\equiv a_\mu n^{*\, \nu}\, .
\ee To the twist-3 accuracy the particle momenta can be written
as: \be{lce} P&=&\frac12(p+p')=n^*, \quad \Delta = p'-p =-2\xi
P+\Delta_\perp, \, \nonumber \\[4mm] q&=&-2\xi
P+\frac{Q^2}{4\xi}n, \quad
q'=q-\Delta=\frac{Q^2}{4\xi}n-\Delta_\perp\, ,
\ee
where $p, p'$
are momenta of the initial and final nucleon and $q,q'$ are
momenta of the initial and final photon respectively.

The DVCS amplitude on the nucleon is related to the matrix element
of T-product of two electromagnetic currents
\be{mat}
T^{ \mu \nu}=-i\int d^4 x e^{-i(q+q')x/2}
\langle p'|T J^\mu(x/2)J^\nu(-x/2)|p \rangle\, .
\ee
To the twist-3 accuracy it has been
computed in Refs.~\ci{Penttinen,BM}. We use here the form of the
amplitude obtained in \cite{Penttinen} in framework of the parton
model \footnote{We change little bit notations of
Ref.~\cite{Penttinen}}. This form is equivalent to that of
Ref.~\cite{BM}, and can be written as:
\be{T}
T^{ \mu \nu}&=&
\frac12 \int_{-1}^1 dx\quad \biggl\{ \left[(-g^{\mu
\nu})_\perp-\frac{P^\nu\Delta_\perp^\mu}{(Pq)}  \right]
 n^\rho F_\rho (x,\xi) C^+(x,\xi)
\nonumber \\[4mm]&-&
\left[(-g^{\nu
k})_\perp-\frac{P^\nu\Delta_\perp^k}{(Pq)}  \right]
i\epsilon^\perp_{k \mu} n^\rho \widetilde F_\rho(x,\xi) C^-(x,\xi)
\nonumber \\[4mm]
&-& \frac{(q+4\xi P)^\mu}{(Pq)} \left[(-g^{\nu
k})_\perp-\frac{P^\nu\Delta_\perp^k}{(Pq)}  \right] \left\{
F_k(x,\xi)C^+(x,\xi)- i\epsilon^\perp_{k \rho}\widetilde F_\rho
(x,\xi) C^-(x,\xi)\right\} \nonumber \\[4mm]&+& \frac{(q+2\xi
P)^\nu}{(Pq)} \left\{ F_{\mu_\perp}(x,\xi)
C^+(x,\xi)+i\epsilon^\perp_{\mu \rho}\widetilde F_\rho (x,\xi)
C^-(x,\xi) \right\}\biggr\}\, , \ee
where the leading order
coefficient functions are
\be{alf} \nonumber
C^\pm(x,\xi)=\frac{1}{x-\xi+i\varepsilon}\pm
\frac{1}{x+\xi-i\varepsilon}\, , \ee
and skewed distributions
$F_\mu(x,\xi)$ and $\widetilde F_\mu(x,\xi)$ are defined in terms
of the nonlocal light-cone quark operators \footnote{The gauge
link between points on the light-cone is not shown but always
assumed.}: \be{Fdef}
F_\mu(x,\xi)&=&\int^\infty_{-\infty}\frac{d\lambda}{2\pi}e^{-ix\lambda} \langle p'|
\pbar(\frac12\lambda n)\gamma_\mu\p (-\frac12\lambda n)
 |p\rangle \, ,
\nonumber \\[4mm] \widetilde
F_\mu(x,\xi)&=&\int^\infty_{-\infty}\frac{d\lambda}{2\pi}e^{-ix\lambda} \langle p'|
\pbar(\frac12\lambda n)\gamma_\mu\gamma_5\p (-\frac12\lambda n)
 |p\rangle \, .
\ee We do not write the sum over quark flavours to simplify
notations, the flavour dependence in Eqs.~\re{Fdef} can  be
restored  easily by the substitution: \be{flav} \pbar \ldots \p
\to \sum_{f=u,d,s, \dots}e_f^2\ \pbar_f \ldots \p_f\, . \ee

The amplitude \re{T} is electromagnetically gauge invariant:
\be{Ginv} q_\mu T^{\mu\nu}=(q-\Delta)_\nu T^{\mu\nu}= 0\, , \ee to
the accuracy $1/Q^2$. In order to have `absolute' transversality
of the amplitude we keep in the expression \re{T} terms of the
$\Delta^2/Q^2$ order, applying the prescription of
\cite{GuichonVander,VGG} ($-g^{\mu\nu}_\perp\to -g^{\mu\nu
}_\perp-\frac{P^\mu\Delta_\perp^\nu}{(Pq)} $) for the twist-3
terms in the amplitude. Formally such terms are beyond our
accuracy and they do not form a complete set of $1/Q^2$
contributions, but we prefer to work with the transverse DVCS
amplitude.

It is easy to see that the third term in Eq.~\re{T} corresponds to
the contribution of the longitudinal polarisation of the virtual
photon. Note that the last term in the expression \re{T},
corresponding to transverse polarisation of the virtual photon, is
proportional to $(q+2\xi P)=q'+\Delta_\perp$. Hence,
 contracting with the transverse polarisation vector $e_\nu(q')$ of
the  final real photon we obtain \be{zero} e_\nu(q')(q+2\xi
P)^\nu=e_\nu(q')\Delta_\perp^\nu\, . \ee
 Therefore, such term  does not contribute to any observable
with the accuracy $O(\Delta/Q)$. The same contribution has been
obtained for the DVCS on the pion. For the case of the pion target
in Ref.~\ci{RW1} it was shown by a direct calculation that the
structure  $(q+2\xi P)$ is the   truncated form of the vector $q'$
and hence such term has zero projection onto the polarisation
vector of the final photon.

In what follows we compute the skewed parton distributions
\re{Fdef} in the so-called Wandzura-Wilczek (WW) approximation.
This means that we neglect the contributions of the quark-gluon
operators, which define what can be called `genuine twist-3'
part.
 We will  demonstrate
that at least in this approximation the amplitude possesses the
factorisation property. Another point, assuming that the
contribution of the quark-gluon correlations us smaller than the WW
part, one can estimate the effects of the $1/Q$ suppressed terms
to various DVCS observables.

\section*{\normalsize \bf Calculation of the matrix elements
in the WW approximation}

The aim of this section is the calculation of the matrix elements
\re{Fdef} in the WW-approximation. To obtain the answer we make
use of the operator identities derived
in~\ci{BM,RW1,KPST}\footnote{For simple derivation
see
Ref.~\cite{KPST}} on the basis of technique developed in ref.~\cite{BB89}:
\be{Oident1} \nonumber \pbar ( x)\gamma_\mu \p
(- x)&=&
 \frac12 \int_0^1 d\alpha
\left\{ \alpha\,   \pert_\mu \left( e^{-  \bar \alpha
(x\pert)}-e^{  \bar \alpha (x\pert)} \right) +\left( e^{-  \bar
\alpha (x\pert)}+e^{  \bar \alpha (x\pert)} \right){\partial \over
\partial x_\mu} \right\} \pbar( \alpha x)\xslash \p(- \alpha x)
\nonumber\\
 &-& i\epsilon_{\mu i j k}  x^i\pert^j
\int_0^1 du\, e^{(2u-1) (x\pert)} \int_0^u d\alpha e^{-\alpha
(x\pert) } { \partial \over \partial x_k}
 \pbar( \alpha x)\xslash\gamma_5  \p (- \alpha x) +\, \dots\, ,
\ee where $\bar \alpha=1-\alpha$ and ellipses stand for the
contributions of either twist-3 quark gluon operators or
twist-4 operators\footnote{ In Ref.~\cite{KPST} more general
relation is derived in which the twist-4 terms of the kind
$x^2\Delta^2$ are summed up.} which we neglect in the WW
approximation. An analogous expression for the operator $\pbar (
x)\gamma_\mu \gamma_5\p (- x)$ can be obtained from
Eq.~\re{Oident1} by the replacement $\xslash\to\xslash\gamma_5$ in
RHS of this equation. Keeping in mind this trivial replacement we
shall present all derivations  only for the vector operator, the
generalisation to axial vector operators is trivial.

The symbol $\pert$ denotes the derivative with respect to the
total translation: \be{Dtot} \pert_{\alpha}\left\{ \pbar(tx)
\Gamma [tx, -tx] \p(-tx) \right\} \equiv \left.
\frac{\partial}{\partial y^{\alpha}} \left\{ \pbar(tx + y) \Gamma
[tx + y, -tx + y] \p(-tx + y)\right\} \right|_{y \rightarrow 0},
\ee with a generic Dirac matrix structure $\Gamma$ and $[x,y]
=\mbox{\rm Pexp}[ig\!\! \int_0^1\!\! dt\,(x-y)_\mu
A^\mu(tx+(1-t)y)] $. Note that in the matrix elements  the total
derivative can be easily converted to  the momentum transfer:
\be{exmpl} \langle p'|\pert_\mu \pbar(tx)\Gamma [tx, -tx] \p(-tx)
|p\rangle=i(p'-p)_\mu \langle p'|\pbar(tx)\Gamma [tx, -tx] \p(-tx)
|p\rangle \ee and therefore in the matrix elements one can
associate $\pert$ with the momentum transfer $\Delta$.
For what follows it is convenient to rewrite the identity
\re{Oident1} in the more symmetric form: \footnote{for simplicity
we use notation $\partial /\partial x_\mu \equiv \partial_\mu$}
\be{Oident2} \pbar ( x)\gamma_\mu \p (- x)&=&
\frac{\pert_\mu}{(\pert x)}\pbar ( x)\xslash \p (- x) \nonumber
\\[4mm]&+& \frac12\int_0^1 d\alpha \left\{ e^{\bar \alpha
(x\pert)}+e^{-\bar \alpha (x\pert)} \right\} \left[
\partial_\mu -\frac{\pert_\mu}{(x\pert)} (x\partial)
\right] \pbar (\alpha x)\xslash \p (-\alpha x) \nonumber
\\[4mm]&+& \frac12\int_0^1 d\alpha \left\{ e^{-\bar \alpha
(x\pert)} - e^{\bar \alpha (x\pert)} \right\} i\epsilon_{\mu
ijk}x_i\frac{\pert_j}{(\pert x)}
\partial_k \pbar (\alpha x)\xslash\gamma_5 \p (-\alpha x)+\dots
\ee To obtain this expression one has to perform an integration by
parts in the first and last terms of Eq.~\re{Oident1}.

Our aim is to calculate the matrix elements $\langle p'| \pbar(
z)\gamma^\mu(\gamma_5)\p (-z) |p\rangle$ in terms of the twist-2
{\it symmetrical} matrix elements $\langle p'|\pbar(
x)\xslash(\gamma_5)\p (-x) |p\rangle$ which appear in the RHS of
Eq.~\re{Oident2}. At the first step, the symmetric matrix elements
should be parametrised for arbitrary vector $x$ which is not
light-like.
The parametrisations of the twist-2 matrix elements can be written
as follows: \be{Otw2} \langle p'| \pbar( x)\xslash\p (-x)
 |p\rangle=
\spin{\xslash}T_1[(Px),\xi(x)]+
\spin{\frac{i\sigma^{\alpha\beta}x_\alpha\Delta_\beta}{2M}}
T_2[(Px),\xi(x)], \ee where we use shorthand notations for:
\be{bispin} \xi(x)=-\frac12 \frac{(\Delta x)}{(Px)},\quad
\spin{\Gamma}\equiv \bar U(p',S')\, \Gamma\,  U(p,S) \, , \ee
where  $U(p,S)$ and $\bar U(p',S')$ denote initial and final nucleon spinors,
respectively. The
amplitudes $T_{1,2}[(Px),\xi(x)]$ can be parametrised in terms of
the twist-2 skewed parton distribution $H,E$ of Ref.~\cite{DVCS2}:
\be{T12} T_1[(Pz),\xi]= \int_{-1}^1 du e^{iu2(Pz)}H(u,\xi),
\nonumber \\[4mm] T_2[(Pz),\xi]= \int_{-1}^1 du
e^{iu2(Pz)}E(u,\xi). \ee Similar formulae can be written for
the axial operator:
\be{Atw2} \langle p'| \pbar( x)\xslash\gamma_5 \p
(-x)
 |p\rangle=
\spin{\xslash\gamma_5}\widetilde T_1[(Px),\xi(x)]+
\spin{\frac{\gamma_5(\Delta x)}{2M}} \widetilde T_2[(Px),\xi(x)],
\ee with amplitudes $\widetilde T_{1,2}$ given in terms of twist-2
SPD's $\widetilde H,\widetilde E$ as: \be{tT12} \widetilde
T_1[(Pz),\xi]&=& \int_{-1}^1 du e^{iu2(Pz)}\widetilde H(u,\xi),
\nonumber \\[4mm] \widetilde T_2[(Pz),\xi]& =& \int_{-1}^1 du
e^{iu2(Pz)}\widetilde E(u,\xi). \ee Now, we have to insert the
parametrisations \re{T12} and \re{tT12} into the RHS of
\re{Oident2} and calculate the matrix elements up to accuracy
$O(\sqrt{x^2})$ which corresponds to the $O(1/Q)$ accuracy in the DVCS
amplitude. To collect contributions with such accuracy we first
keep the vector $x$ off light-cone and after performing the
differentiation with respect to vector $x$ we assign to the vector
$x$ the value on the light-cone $x\rightarrow z,\ z^2=0$.

Let us illustrate the main steps of the calculations on the
example of matrix element of the vector operator. The first term
in the RHS of the \re{Oident2} can easily be  computed with the
result: \be{1th} \langle p'| \frac{\pert^\mu}{(\pert x)}\pbar (
x)\xslash \p (- x)
 |p\rangle &=&- \frac{\Delta^\mu}{2\xi}\spin{\gamma_+}
(T_1[Pz,\xi]+T_2[Pz,\xi])+ \nonumber \\[4mm]
&+&\frac{\Delta^\mu}{2\xi}\spin{\frac 1 M}T_2[Pz,\xi]\, +\, ...\,
, \ee where ellipses stand for the twist-4 and higher corrections.
To get the above expression we used that \be{gplus} \gamma_\mu
z^\mu/(Pz)=\gamma_\mu n^\mu\equiv \gamma_+ \, , \ee and the Gordon
identity: \be{si1}
\spin{\frac{i\sigma^{\alpha\beta}x_\alpha\Delta_\beta}{2M}}=
-(Px)\spin{\frac 1 M}+\spin{\xslash}, \ee which  follows from the
equations of motion for nucleon Dirac spinors: \be{EOM}
(\pslash-M)U(p,S)=0,\quad \bar U(p',S')(\pslash'-M)=0\, . \ee

Consider now a more complicated matrix element of the second term
in RHS of \re{Oident2}. Using \re{Dtot}, \re{Otw2} and \re{si1} we
obtain for this term the following expression:
\be{2th1} 2^{\rm nd}\, {\rm term}& =& \frac12\int_0^1 d\alpha \left\{
e^{-i\bar\alpha(\Delta x)}+e^{i\bar\alpha(\Delta x)} \right\}
[\partial_\mu-\frac{\Delta_\mu}{(\Delta x)}(x\partial)]\ \nonumber\\[4mm]
&\times& \biggl( \spin{\xslash}(
T_1[(Px),\xi(x)]+T_2[(Px),\xi(x)])
-\spin{\frac{(Px)}{M}}T_2[\alpha (Px),\xi(x)]\biggl)\, . \ee
Note,
we can put $x=z$ only after differentiation! The latter can be
done with the help of the following formulae:
\be{deriv1}
\partial_\mu T_i[\alpha (Px),\xi(x)]|_{x=z}=
\frac1{(Pz)}\biggl[ P^\mu \alpha\frac{\partial}{\partial \alpha}-
\frac12\Delta_\perp^\mu \frac{\partial}{\partial \xi} \biggl]
T_i[\alpha (Pz),\xi]\, , \ee \be{deriv2} [\partial_\mu-
\frac{\Delta_\mu}{(\Delta x)}(x\partial)]T_i[\alpha (Px),\xi(x)]|_{x=z}=
\frac1{(Pz)}\frac{\Delta_\perp^\mu}{2\xi} \biggl(
\alpha\frac{\partial}{\partial \alpha}
-\xi\frac{\partial}{\partial \xi} \biggl)T_i[\alpha (Pz),\xi]\, .
\ee Applying these formulae and using simple identity:
\be{sident2} P^\mu \spin{\zslash}=
(Pz)\spin{\gamma^\mu}-(Pz)\spin{\gamma^\mu_\perp}\, , \ee we
obtain for the second term  the following expression: \be{2th2}
2^{\rm nd}\, {\rm term} &=& \frac12\int_0^1 d\alpha \left\{
e^{i\bar\alpha 2\xi (Px)}+e^{-i\bar\alpha 2\xi(Px)}
\right\}\biggl( \spin{\gamma^\mu_\perp}( T_1+T_2)[\alpha(Pz),\xi]+
\nonumber\\[4mm]&& +\frac{\Delta_\perp^\mu}{2\xi}\spin{\gamma_+}
\biggl(1+\alpha\frac{\partial}{\partial \alpha}-
\xi\frac{\partial}{\partial \xi}\biggl) (
T_1+T_2)[\alpha(Pz),\xi]- \nonumber\\[4mm]&&
-\frac{\Delta_\perp^\mu}{2\xi}\spin{\frac 1M}
\biggl(1+\alpha\frac{\partial}{\partial \alpha}-
\xi\frac{\partial}{\partial \xi}\biggl) T_2[\alpha(Pz),\xi]\, .
\ee Using the definition of the amplitudes $T_{1,2}$ Eqs.~\re{T12}
this term can be represented in the form which is convenient for
the
transition to the momentum space: \be{2th3} 2^{\rm nd}\, {\rm
term} =\int_{-1}^{1}du\ G^\mu(u,\xi) \int_0^1 d\alpha\left\{
e^{i2(Px)[\alpha(u- \xi)+\xi]}+e^{i2(Px)[\alpha(u+\xi)-\xi]}
\right\} \, , \ee where \be{G} G^\mu(u,\xi)&=&
\spin{\gamma^\mu_\perp}(H+E)(u,\xi)+ \frac{\Delta_\perp^\mu}{2\xi}
\spin{\frac 1 M}\biggl[u\frac{\partial}{\partial u}+
\xi\frac{\partial}{\partial \xi} \biggl] E(u,\xi)-
\nonumber\\[4mm]&& -\frac{\Delta_\perp^\mu}{2\xi}
\spin{\gamma_+}\biggl[u\frac{\partial}{\partial u}+
\xi\frac{\partial}{\partial \xi}\biggl] (H+E)(u,\xi)\, . \ee
Manipulations with the nucleon matrix element of the third term of
the operator identity~\re{Oident2} are similar. For this term
 we obtain:
\be{3th} 3^{\rm d}\, {\rm term} &=& i\epsilon^\perp_{\mu k}
\int_{-1}^{1}du\ \widetilde G^k(u,\xi) \int_0^1 d\alpha\left\{
e^{i2(Px)[\alpha(u- \xi)+\xi]}-e^{i2(Px)[\alpha(u+\xi)-\xi]}
\right\} \, , \ee where \be{tG} \widetilde G^k(u,\xi)& =
&\spin{\gamma^\mu_\perp\gamma_5} \widetilde H(u,\xi)
+\frac12\Delta_\perp^\mu \spin{\frac{\gamma_5}{M}}
\biggl[1+u\frac{\partial}{\partial u}+\xi\frac{\partial}{\partial
\xi}\biggl] \widetilde E(u,\xi)- \nonumber\\[4mm]&&
-\frac{\Delta_\perp^\mu}{2\xi}\spin{\gamma_+\gamma_5}
\biggl[u\frac{\partial}{\partial u}+\xi\frac{\partial}{\partial
\xi}\biggl] \widetilde H(u,\xi) \, . \ee Collecting all terms
\re{1th},\re{2th3} and \re{3th} and performing the Fourier
transformation on the light-cone we obtain the expression for
$F_\mu$ (see definition~\re{Fdef}) in the WW approximation in
terms of twist-2 SPD's $H,E$ and $\widetilde H, \widetilde E$:
\be{F} F^{WW}_\mu(x,\xi)&=&  \frac{\Delta_\mu}{2\xi}\spin{\frac 1
M}E(x,\xi)- \frac{\Delta_\mu}{2\xi}\spin{\gamma_+}(H+E)(x,\xi)+
\nonumber\\[4mm]&&\mskip-10mu +\int_{-1}^{1}du\
G_\mu(u,\xi)W_{+}(x,u,\xi)+i\epsilon_{\perp \mu k}
\int_{-1}^{1}du\  \widetilde G^k (u,\xi)W_{-}(x,u,\xi)\, , \ee
where we introduced a convenient notation for the WW-kernels:
\be{Wpm} W_{\pm}(x,u,\xi)&=& \frac12\biggl\{
\theta(x>\xi)\frac{\theta(u>x)}{u-\xi}-
\theta(x<\xi)\frac{\theta(u<x)}{u-\xi} \biggl\} \nonumber
\\[4mm]&&\mskip-10mu \pm\frac12\biggl\{
\theta(x>-\xi)\frac{\theta(u>x)}{u+\xi}-
\theta(x<-\xi)\frac{\theta(u<x)}{u+\xi} \biggl\}. \ee The function
$G^\mu(u,\xi)$ and $\widetilde G^k (u,\xi)$ are defined in \re{G}
and \re{tG}.

The analogous calculation  for the axial matrix elements gives:
\be{F5} \widetilde F^{WW}_\mu(x,\xi)&=&  \frac{\Delta_\mu}{2}
\spin{\frac{\gamma_5}{M}}\widetilde E(x,\xi)-
\frac{\Delta_\mu}{2\xi}\spin{\gamma_+\gamma_5}\widetilde H(x,\xi)+
\nonumber\\[4mm]&& +\int_{-1}^{1}du\ \widetilde
G_\mu(u,\xi)W_{+}(x,u,\xi)+i \epsilon_{\perp \mu k}
\int_{-1}^{1}du\  G^k (u,\xi)W_{-}(x,u,\xi)\, . \ee
 Let us note that the spinor structures in the functions $G^\mu(u,\xi)$
 and $\widetilde G^k (u,\xi)$ are {\it not} independent.
They can be related to each other with the help of equations of motion
for Dirac spinors~\re{EOM}. However, we prefer not to reduce the
number of structures to retain the symmetric form of the results
(\ref{F},\ref{F5}).

The generalisation of the WW relations for the non-forward case
was discussed first in Ref.~\cite{BlRo}. Our results
(\ref{F},\ref{F5}) do not agree, however, with those of
Ref.~\cite{BlRo}. The reason for the disagreement is that in
Ref.~\cite{BlRo} instead of the complete operator identity of the type
\re{Oident1} only its part which survives in the forward limit has
been considered. This simplification is justified  in order  to
reproduce only the WW relations in the forward case. The WW
relations obtained in Ref.~\cite{BM} can be reduced to the form
(\ref{F},\ref{F5}) obtained here.

\section*{\normalsize \bf Properties of the WW kernels }

In the previous section we introduced the notion of the
WW-kernels, see Eq.~\re{Wpm}. In this section we consider the
general properties of these kernels. Let us introduce the
following notations for the action of the WW kernels on a function
$f(u,\xi)$:

\be{conv} W_\pm\otimes f [x,\xi]\equiv \int_{-1}^1du\
W_\pm(x,u,\xi) f(u,\xi)\, , \ee with $W_\pm$ given by
Eq.~\re{Wpm}. We shall call ``WW transform'' the resulting
functions $W_\pm\otimes f [x,\xi]$ and we shall call the ``WW
transformation'' the action of the WW kernels.

\noindent {\it Limiting cases}.
 We consider two limiting cases of the WW transformation: the forward
limit $\xi\to 0$ and the `meson' limit $\xi\to 1$. In the forward
limit we easily obtain: \be{fwd} \nonumber \lim_{\xi\to 0}
W_+\otimes f [x,\xi]&=& \theta(x\ge 0)\int_x^1 \frac{du}{u}\
f(u,\xi=0)- \theta(x\le 0)\int_{-1}^x \frac{du}{u}\ f(u,\xi=0)\,
,\\ \lim_{\xi\to 0} W_-\otimes f [x,\xi]&=&0\, . \ee We can see that
the action of the $W_+$ in the forward limit reproduces the
Wandzura--Wilczek relation for the spin structure function $g_T$
\cite{WW}. The term with $\theta(x\ge 0)$ corresponds to the quark
distributions and the term with $\theta(x\le 0)$ to the antiquark
ones. The $W_-$ kernel disappears in the forward limit, so that
this kernel is `genuine non-forward' object.

In the limit $\xi\to 1$ the skewed parton distributions have
properties of meson distribution amplitudes. In this limit the WW
transforms have the form (we use notation
$f(u,\xi=1)=\varphi(u)$): \be{meson} \nonumber \lim_{\xi\to 1}
W_\pm\otimes f [x,\xi]&=&\frac 12 \biggl\{ \int_{-1}^x
\frac{du}{1-u}\ \varphi(u)\pm \int_{x}^1 \frac{du}{1+u}\
\varphi(u)\biggr\}\, , \ee which corresponds to the WW relations
for the meson distribution amplitudes derived in \cite{BB,BBKT}.

We can see that WW transforms of skewed parton distributions
interpolate between WW relations for parton distributions and the
meson distribution amplitude. Also general form of the WW kernels
\re{Wpm} allows to derive WW relations for distribution amplitudes
of a meson of arbitrary spin.

\noindent {\it Mellin moments}: One can easily derive the Mellin
moments of the WW transform. The result has the form:

\be{Mellin} \int_{-1}^1dx \ x^N\ W_\pm\otimes f
[x,\xi]=\frac{1}{N+1} \int_{-1}^1 du \left[
\frac{u^{N+1}-\xi^{N+1}}{u-\xi}\pm
 \frac{u^{N+1}-(-\xi)^{N+1}}{u+\xi}\right] f(u,\xi)\, .
\ee From this simple exercise we can see important property of the WW
transformation, namely, if the function $f(u,\xi)$ satisfies the
polynomiality condition, {\it i.e.}:

\be{poly} \int_{-1}^1du\ u^N\ f(u,\xi)={\rm polynomial\ in\ } \xi\
{\rm of\ the\ order\ } N+1\, , \ee its WW transform also satisfies
the polynomiality condition. Let us consider now the special cases
$N=0,1$ of the  Mellin moments. These particular moments of the
twist-3 SPD's \re{Fdef} do not receive contribution from the
`genuine twist-3' quark-gluon operators, therefore the WW
approximation for them gives exact results. In the forward limit
this observation leads to Burkhard-Cottingham ($N=0$) \cite{BC}
and Efremov-Leader-Teryaev ($N=1$) \cite{ELT} sum rules for
polarized structure function $g_T$. The corresponding
generalizations of these sum rules in the case of skewed parton
distributions have been discussed in Refs.~\cite{Penttinen,KPST}.
From general expression \re{Mellin} one gets: \be{n01} \nonumber
\int_{-1}^1 dx\ W_-\otimes f [x,\xi]&=&0\, ,\\ \nonumber
\int_{-1}^1 dx\ W_+\otimes f [x,\xi]&=&\int_{-1}^1 du\ f(u,\xi)\,
,\\ \int_{-1}^1 dx\ x\ W_-\otimes f [x,\xi]&=& \frac\xi 2
\int_{-1}^1 du\ f(u,\xi)\, ,\\ \nonumber \int_{-1}^1 dx\ x\
W_+\otimes f [x,\xi]&=& \frac 1 2 \int_{-1}^1 du\ u\ f(u,\xi)\, .
\ee Below we shall use these results for the consistency checks of
our results for skewed parton distributions and for derivation of
the sum rules for the distributions $F_\mu$ and $\widetilde F_\mu$
\re{Fdef}.\\ {\it Discontinuities}. In Ref.~\cite{KPST} it was
demonstrated that the twist-3 skewed parton distributions in the
WW approximation exhibit discontinuities at the points
$x=\pm\xi$. This feature is related to the properties of the WW
kernels. Let us compute the discontinuities of a WW transform at
the points $x=\pm \xi$: \be{skachki} \nonumber \lim_{\delta\to 0}
\left[ W_\pm\otimes f [\xi+\delta,\xi]-W_\pm\otimes f
[\xi-\delta,\xi] \right]&=&\frac 12 vp\int_{-1}^1 \frac{du}{u-\xi}
f(u,\xi)\, ,\\ \lim_{\delta\to 0} \left[ W_\pm\otimes f
[-\xi+\delta,\xi]-W_\pm\otimes f [-\xi-\delta,\xi]
\right]&=&\pm\frac 12 vp\int_{-1}^1 \frac{du}{u+\xi} f(u,\xi)\, .
\ee Here $vp$ means an integral in the sense of {\it valeur
principal}. We see that for a very wide class of functions
$f(u,\xi)$ the discontinuity of the corresponding WW transforms
is nonzero. This feature of the WW transformation may lead to the
violation of the factorisation for  the twist-3 DVCS amplitude.

\section*{\normalsize \bf Properties of the skewed distributions
$F_\mu$ and $\widetilde F_\mu$}

In this section we analyse the general properties of the skewed
parton distributions $F^{WW}_\mu$ and $\widetilde F^{WW}_\mu$
\re{Fdef} in the WW approximation. We consider the limiting cases
and derive sum rules for these SPD's which are valid beyond WW
approximation.

\noindent {\it Forward limit}: In the forward limit the `vector'
SPD $F_\mu$ in the WW approximation \re{F} vanishes, because
 unpolarized parton densities  of twist-3 are absent.
The forward limit for the `axial' SPD $\widetilde
F_{\mu_\perp}^{WW}$, see \re{F5}, is nontrivial: \be{fwd5}
\nonumber \lim_{\Delta\to 0} \widetilde F_{\mu_\perp}^{WW}(x,\xi)=
2 S_\mu^\perp \biggl[ \Delta q(x) +\int_x^1 \frac{du}{u} \Delta
q(u)\biggr] \, \quad {\rm for}\ x\ge 0 \, ,\\ \lim_{\Delta\to 0}
\widetilde F_{\mu_\perp}^{WW}(x,\xi)= 2 S_\mu^\perp \biggl[ \Delta
\bar q(|x|) +\int_{|x|}^1 \frac{du}{u} \Delta \bar q(u)\biggr] \,
\quad {\rm for}\ x\le 0 \, . \ee It is easy to recognize in these
equations the Wandzura--Wilczek relations for $g_T(x)$ \cite{WW}.
In the derivation \re{fwd5} we used  that twist-2 skewed
distribution $\widetilde H$ is reduced in the forward limit to the
polarized quark (for $x\ge 0$) and antiquark (for $x\le 0$)
densities $\Delta q(x)$ and $\Delta \bar q(x)$.

\noindent {\it Mellin moments}: The lowest $N=0,1$ Mellin moments
of the SPD's $F_\mu$ and $\widetilde F_\mu$ \re{Fdef} have been
analysed previously in Ref.~\cite{Penttinen}.
 Since  for these moments
the contributions from the twist-3 quark-gluon operators are
absent,  we have to reproduce sum rules of Ref.~\cite{Penttinen}
from the  WW approximation for the functions $ F_\mu$ and
$\widetilde F_\mu$, see Eqs.~\re{F},\re{F5}. This would be a
powerful check of our results.

{} As it follows from the definitions of twist-3 SPD's \re{Fdef},
 the $N=0$  moments are reduced to the nucleon matrix elements
of the local currents which are parametrized in terms of electric,
magnetic,  axial and pseudoscalar form factors of the nucleon:
\be{M0} \int_{-1}^1du\
F_\mu(u,\xi)&=&\spin{\gamma_\mu}F_1(\Delta^2)+
\spin{\frac{i\sigma_{\mu\nu}\Delta^\nu}{2M}}F_2(\Delta^2)\, ,\\
\int_{-1}^1du\ \widetilde F_\mu(u,\xi)&=&
\spin{\gamma_\mu\gamma_5}G_A(\Delta^2)+
\frac{\Delta_\mu}{2M}\spin{\gamma_5}G_P(\Delta^2)\, . \ee Using
the properties of the WW kernels \re{n01} one can easily see that
the expressions for $F_\mu^{WW}$ \re{F} and for $\widetilde
F_\mu^{WW}$ \re{F5} satisfy the general QCD sum rules \re{M0}.
This is a sensitive check of our approach.

The general QCD expression for the $N=1$  moment of the SPD's
$F_\mu$ and $\widetilde F_\mu$ can be obtained from the WW
approximation because the quark-gluon operators do not contribute
to this moment. Again with help of \re{n01} we can obtain the
results for the $N=1$
 moments of the transverse components of the functions
$F_\mu^{WW}$ and $\widetilde F_\mu^{WW}$ which enter the
 DVCS amplitude for the longitudinally polarized virtual photon.
For the function $F_{\mu_\perp}$ we have
\be{M1v} \nonumber
\int_{-1}^1du\ u\
F_{\mu_\perp}^{WW}(u,\xi)&=&\spin{\gamma^\perp_\mu}\ \frac {1}2
\left[ \int_{-1}^1du\ u
\left\{H(u,\xi)+E(u,\xi)\right\}+G_A(\Delta^2)\right]\\
&+&\frac{\Delta^\perp_\mu}{4M}\spin{1} \frac{\partial}{\partial
\xi}\ \int_{-1}^1du\ u \ E(u,\xi)\, . \ee We see that the
`transverse' SPD $F_{\mu_\perp}$ is sensitive to the combination
$$\int_{-1}^1du\ u \left(H(u,\xi)+E(u,\xi)\right)+G_A(\Delta^2)$$
which is related to the spin structure of the nucleon in the
forward  limit: \be{spin} \lim_{\Delta\to 0}\int_{-1}^1du\ u
\left\{H(u,\xi)+E(u,\xi)\right\} + G_A(\Delta^2)= 2 J_q + \Delta
q\, , \ee where $J_q$ is a fraction of the total angular momentum
of the nucleon carried by quarks and $\Delta q$ is the
corresponding fraction of the spin. To derive Eq.~\re{spin} we
made use of Ji's sum rule \cite{DVCS2}. Also the new angular
momentum sum rule \re{M1v} coincides with that previously derived
in \cite{Penttinen} where the  WW approximation has not been
considered.
 We see that the DVCS amplitude with longitudinally polarized
virtual photon is also sensitive to the spin structure of the
nucleon. In principle, this part of the amplitude can be
extracted from the data through the angular, spin and $Q$
dependence of the differential cross section
\cite{Diehl,Belitsky:2000gz}. This may provide us with additional
possibility to probe $J_q$ in the nucleon.

In the case of axial transverse  SPD $\widetilde F_{\mu_\perp}$ we
obtain from Eq.~\re{F5} the following result for the $N=1$ moment:
\be{M1a} \nonumber \int_{-1}^1du\ u\ \widetilde
F_{\mu_\perp}^{WW}(u,\xi)&=&\spin{\gamma^\perp_\mu \gamma_5}\
\frac {1}2 \left[ \int_{-1}^1du\ u\ \widetilde H(u,\xi)+ \xi^2
\{F_1(\Delta^2)+ F_2(\Delta^2)\}\right]\\&+& {\rm other\
structures} \ee For simplicity we do not write all possible
spinor structures which appear in the RHS \re{M1a}. For the
comparison with the sum rules of Ref.~\cite{Penttinen} we need
only coefficient in front of structure
$\spin{\gamma^\mu_\perp\gamma_5}$. We obtained that our result
\re{M1a} is in agreement with Ref.~\cite{Penttinen}. In the
forward limit we reproduce from the Eq.~\re{M1a} the
Efremov-Leader-Teryaev sum rule \cite{ELT} for the polarized
structure function $g_T(x)$.

\noindent {\it Discontinuities}:
Using general properties of the WW transformation \re{skachki} and
Eqs. \re{F} and \re{F5} one obtains that $F_\mu^{WW}$ and
$\widetilde F_\mu^{WW}$ have discontinuities at the points $x=\pm
\xi$ . But using a certain  symmetry of the equations \re{F} and
\re{F5}, one can find that some combinations of the distributions
$F_\mu^{WW}$ and $\widetilde F_\mu^{WW}$  are free of
discontinuities. For example, using  formulae~\re{skachki} one can
see that the combination:

\be{comb1} F_\mu^{WW}(x,\xi)-i\varepsilon_{\perp\mu\rho}
\widetilde F_\rho^{WW}(x,\xi)\, , \ee has no discontinuity at
$x=\xi$. On the other hand, the  `dual' combination: \be{comb2}
F_\mu^{WW}(x,\xi)+ i\varepsilon_{\perp\mu\rho} \widetilde
F_\rho^{WW}(x,\xi)\, , \ee is free of the discontinuity at
$x=-\xi$. As we shall see below,  the cancellation of
discontinuities in these particular combinations of the SPD's
ensures the factorisation of the twist-3 DVCS amplitude on the
nucleon.

\subsection*{\normalsize \bf  Pion pole and D-term in $F_\mu$
and $\widetilde F_\mu$} In the Wandzura-Wilczek approximation the
new skewed distributions $F_\mu$ and $\widetilde F_\mu$ (see
Eq.~\re{Fdef} for definition) are fixed completely in terms of the
twist-2 SPD's $H,E$ and $\widetilde H, \widetilde E$. These
functions have some specific contributions which are invisible in
the forward limit. In this section we consider how these special
contributions are incorporated into the structure of the twist-3
distributions.

\noindent {\it Pion pole contribution:} The skewed parton
distribution $\widetilde E$ at small $t$ is dominated by chiral
contribution of the pion pole \cite{pionpole,pp2,pp3,pp4} of the form:
\be{pipo} \widetilde E^{\rm pion\ pole}(x,\xi)=\frac{4 g_A^2
M^2}{-t+m_\pi^2}\ \frac 1\xi \varphi_\pi\left(\frac x\xi
\right)\theta(|x|\le\xi)\, , \ee where $g_A$ is the axial charge
of the nucleon and $\varphi_\pi(u)$ is the pion distribution
amplitude. In the WW relation (\ref{F}) the twist-2 SPD
$\widetilde E$ enters only in the combination: \be{killpipo}
\left[ 1+ u\frac{\partial}{\partial
u}+\xi\frac{\partial}{\partial\xi} \right] \widetilde E(u,\xi)\, .
\ee
 One can easily see that the contribution of the pion pole \re{pipo}
nullifies under action of the differential operator in
Eq.~\re{killpipo}. The only place where the pion pole contribution
survives is the SPD $\widetilde F_\mu$, see Eq.~\re{F5}. In this
way we come to the following simple results for the contribution
of the pion pole to the twist-3 SPD's in the WW approximation:
\be{pipo3} \nonumber F_\mu^{WW,{\rm pion\ pole}}(x,\xi)&=&0\, ,\\
\widetilde F_\mu^{WW,{\rm pion\ pole}}(x,\xi)&=&
\frac{\Delta_\mu}{2M}\spin{\gamma_5}\frac{4 g_A^2
M^2}{-t+m_\pi^2}\ \frac 1\xi \varphi_\pi\left(\frac x\xi
\right)\theta(|x|\le\xi)\, . \ee We see an interesting result that
the pion pole contribution to the twist-3 SPD's is expressed in
terms of twist-2 pion distributions amplitude. This observation is
in nice agreement with the fact that in the WW approximation there
exists no twist-3 pion distribution amplitude associated with
vector- and axial-vector operators \cite{BF,PB}.

\noindent {\it D-term contribution:} Now we discuss a specific
contribution to the twist-2 SPD's $H$ and $E$ which was found in
Ref.~\cite{PW}. This contribution completes the parametrisation of
SPD's in terms of double distributions \cite{RadDD}. This
contribution has the form: \be{Dterm} \nonumber H^{\rm
D-term}(x,\xi)&=&~D\left(\frac{x}{\xi}\right)\ \theta(|x|\le\xi)\,
,\\ E^{\rm D-term}(x,\xi)&=&-D\left(\frac{x}{\xi}\right)\
\theta(|x|\le\xi)\, , \ee where the function $D(u)$ is odd in its
argument. The specifics of the D-term contribution is that it is not
`generated' by double distributions. Numerically it can be rather
large as it follows from estimates in the chiral quark-soliton
model of the nucleon \cite{bor}. The importance of the D-term is
especially emphasized by the fact that it contributes considerably
to the DVCS amplitude but drops out in Ji's angular momentum sum
rule \cite{DVCS2} because in the combination $H+E$ D-term cancels.
Now if we substitue the D-term contributions \re{Dterm} to our WW
relations (\ref{F},\ref{F5}) we see that the result is:
\be{Dterm3} \nonumber F_\mu^{\rm WW\
D-term}(x,\xi)&=&-\frac{\Delta_\mu}{2\xi\ M}\spin{1}
D\left(\frac{x}{\xi}\right)\ \theta(|x|\le\xi)\, ,\\ \widetilde
F_\mu^{\rm WW\ D-term}(x,\xi)&=&0\, . \ee Again we reproduced the
result anticipated in Ref.~\cite{Penttinen}.

\section*{\normalsize \bf DVCS amplitude}

Here we analyse the general structure of the DVCS amplitude on the
nucleon to the twist-3 accuracy. To this accuracy the amplitude
receives contributions not only from the transverse polarization of
the virtual photon (as it is at the twist-2 level), but also from
the longitudinal one.

The amplitude with the transverse polarization of the virtual
photon has the form\footnote{We neglected the last term in the
expression \re{T} for the amplitude as it being contracted with
polarization of the final real photon contributes only to the
order $O(1/Q^2)$ which is beyond our accuracy} (see Eq.~\re{T})

\be{Tt} e_\mu^\perp (q) T^{ \mu \nu}&=& \frac12 \int_{-1}^1 dx\
e^\perp_\mu(q)\  \biggl\{
\left[(-g^{\mu\nu})_\perp-\frac{P^\nu\Delta_\perp^\mu}{(Pq)}
\right] F_+(x,\xi) C^+(x,\xi) \nonumber \\[4mm]&-& \left[(-g^{\nu
k})_\perp-\frac{P^\nu\Delta_\perp^k}{(Pq)}  \right]
i\epsilon^\perp_{k \mu}\widetilde F_+(x,\xi) C^-(x,\xi)\biggr\} \,. \ee
We see that in this case the amplitude to the $1/Q$ accuracy
is fixed completely in terms of the same combinations of the
twist-2 SPD's.
Such result for the DVCS  amplitude to the $1/Q$ accuracy confirms
the {\it ad hoc} prescription suggested in
Ref.~\cite{GuichonVander,VGG}. However the twist-3 DVCS amplitude also
receives contribution from the longitudinal polarization of the
virtual photon (see Eq.~\re{T}):

\be{TL} \nonumber e^L_\mu(q)T^{ \mu \nu}&\propto& \left[(-g^{\nu
k})_\perp-\frac{P^\nu\Delta_\perp^k}{(Pq)}  \right]\\
&\times&\int_{-1}^1 dx \left\{ F_k(x,\xi)C^+(x,\xi)-
i\epsilon_{\perp k \rho}\widetilde F_\rho (x,\xi)
C^-(x,\xi)\right\}\, . \ee This part of the amplitude depends on
the `transverse' parts of SPD's $F_{\mu_\perp}$ and $\widetilde
F_{\mu_\perp}$ which are related to twist-2 SPD's $H,E$ and
$\widetilde H, \widetilde E$ {\it only} in Wandzura-Wilczek
approximation. Since the coefficient functions $C^{\pm}(x,\xi)$
have singularities at points $x=\pm \xi$ the integral over $x$ may
diverge due to discontinuities of SPD's in WW approximation.
However it is easy to see that near the point $x=\xi$ SPD's enter
in the combination~\re{comb1} and near the point $x=-\xi$ in the
combination~\re{comb2}. These are exactly the combinations which
are free of discontinuities at the corresponding point, therefore
the amplitude \re{TL} is well defined in the WW approximation. The
same cancellation of the discontinuities was observed in the case
of DVCS on the pion target \cite{RW1,KPST}.  In Ref.~\cite{KPST}
some general arguments why this phenomenon should persist to the
case of DVCS on any target was given. Here we have checked by the
explicit calculations this conjecture for the case of the nucleon
target.

\section*{\normalsize \bf Conclusions}

We have presented a simple and detailed derivation of the
Wandzura-Wilczek relations for the twist-3 skewed parton
distributions in the nucleon. These relations allow to express the
DVCS amplitude on the nucleon to the twist-3 accuracy ($1/Q$) in
terms of the twist-2 SPD's $H,E$ and $\widetilde H, \widetilde E$.

The Wandzura-Wilczek approximation corresponds to neglecting
the nucleon matrix elements of the twist-3 quark-gluon operators.
It would be very interesting to check in various models of the
nucleon structure whether the neglected matrix elements are indeed
small. The theory of instanton vacuum predicts that in the case of
polarized structure functions $g_T(x)$ and $h_L$ the corresponding
matrix elements are parametrically small in the packing fraction
of the instantons in the QCD vacuum, for details see
\cite{BPW,DP}. Another interesting possibility to check the
hypothesis behind the WW approximation is the model where SPD's
are represented as an overlap of nucleon wave functions
\cite{overlap}. The quark-gluon correlations appear in this case
as the overlap between the lowest Fock component of the nucleon
wave function and  that with an additional gluon.

We have checked by  an explicit calculation that although the SPD's
in the WW approximation have discontinuities, the corresponding
would-be divergences in the twist-3 DVCS amplitude are cancelled
exactly and do not lead to problems with the factorization. Of
course, this is not a  proof of the factorization for the twist-3
DVCS amplitude.

We rederive the new angular momentum sum rule of
Ref.~\cite{Penttinen} which allows us to relate twist-3 SPD's to
the angular momentum fraction carried by quarks in the nucleon.
Note that this sum rule is valid beyond the WW approximation.
Since, in principle, the twist-3 SPD's can be extracted from the
experimental data by considering  $Q$, spin and angular
dependencies of the DVCS observables, this opens a complementary
approach to the measurement of the quark angular momentum
distribution, compared to the original proposal by Ji \ci{DVCS2}.

The twist-3 contributions are suppressed in the DVCS amplitude
relative to the twist-2 one by one power of $1/Q$  only. Therefore,
 estimates of these corrections are extremely important for the
possibility to extract twist-2 SPD's from experimental data. The
size of the twist-3 corrections increases with the momentum
transfer squared $t$, because these corrections typically enter
the amplitude as $\sqrt{-t}/Q$. Their studies are hence, also
helpful for the analysis of the $t-$dependence of twist-2 SPD. The
latter task is mandatory for the very possibility to extract
information about $L_q$ from the data because $L_q$ is related to
the SPD's only at $t=0$.  This point lies outside the physical
region of the DVCS process so that extrapolation is required.

\section*{\normalsize \bf Acknowledgments.}
We would like to thank  I.~Anikin, V.~Braun, A.~Belitsky,
L.~Frankfurt, K.~Goeke, L.~Mankiewicz, D.~M\"uller, M.~Penttinen,
P.~Pobylitsa, A.~Radyushkin, A.~Sch\"afer, M.~Strikman, M.~Vanderhaeghen, and
C.~Weiss for fruitful discussions.
M.V.P. acknowledges kind
hospitality of BNL and Penn State University where a part of this
work has been done. The work of N.K. was supported by the DFG,
project No.~920585. M.V.P. is supported by DFG, BMFB and COSY.

\section*{\normalsize \bf APPENDIX: }
\setcounter{equation}{0} \label{app:a}
\renewcommand{\theequation}{A.\arabic{equation}}
\setcounter{table}{0}
\renewcommand{\thetable}{\Alph{table}}

For the convenience of the reader who may wish to use our results
for the calculation of DVCS observables to the twist-3 accuracy,
we collect the resulting expressions for the DVCS amplitude to the
accuracy $O(1/Q)$ in the Wandzura-Wilczek approximation. We
present the results in the form which can be easily used for the
analysis of DVCS observables.

The DVCS amplitude on the nucleon to the order $O(1/Q)$ obtained
in Refs.~\cite{Penttinen,BM} has the form: \be{Tapp} T^{ \mu
\nu}&=& \frac12 \int_{-1}^1 dx\quad \biggl\{ \left[(-g^{\mu
\nu})_\perp-\frac{P^\nu\Delta_\perp^\mu}{(Pq)}  \right] n^\beta
F_\beta (x,\xi) C^+(x,\xi) \nonumber \\[4mm]&-& \left[(-g^{\nu
k})_\perp-\frac{P^\nu\Delta_\perp^k}{(Pq)}  \right]
i\epsilon^\perp_{k \mu} n^\beta \widetilde F_\beta (x,\xi)
C^-(x,\xi) \nonumber \\[4mm]&-& \frac{(q+4\xi P)^\mu}{(Pq)}
\left[(-g^{\nu k})_\perp-\frac{P^\nu\Delta_\perp^k}{(Pq)}  \right]
\left\{ F_k(x,\xi)C^+(x,\xi)- i\epsilon^\perp_{k \rho}\widetilde
F_\rho (x,\xi) C^-(x,\xi)\right\} \biggr\} \nonumber \ee where to
the twist-3 accuracy: \be{lceapp} P&=&\frac12(p+p')=n^*, \quad
\Delta = p'-p =-2\xi P+\Delta_\perp, \, \nonumber \\[4mm]
q&=&-2\xi P+\frac{Q^2}{4\xi}n, \quad
q'=q-\Delta=\frac{Q^2}{4\xi}n-\Delta_\perp \ee where $p, p'$ are
momenta of the initial and final nucleon and $q,q'$ are momenta of
the initial and final photon respectively. The light-like vectors
$n$ and $n^*$ are normalized as $(n\cdot n^*)=1$. Also we
introduce the metric and totally antisymmetric tensors in the two
dimensional transverse plane ($\varepsilon_{0123}=+1$):

\be{gtapp} (-g^{\mu \nu})_\perp = -g^{\mu \nu}+ n^\mu n^{*
\nu}+n^\nu n^{* \mu}, \quad \epsilon^\perp_{\mu \nu}=
\epsilon_{\mu \nu \alpha\beta}n^\alpha P^{\beta} \ee The leading
order coefficient functions are: \be{alfapp} \nonumber
C^\pm(x,\xi)=\frac{1}{x-\xi+i\varepsilon}\pm
\frac{1}{x+\xi-i\varepsilon}. \ee The skewed parton distributions
$F_\mu$ and $\widetilde F_\mu$ can be related to the twist-2 SPD's
$H,E,\widetilde H$ and $\widetilde E$ with help of
Wandzura-Wilczek relations:

\be{Fapp} F^{WW}_\mu(x,\xi)&=&  \frac{\Delta_\mu}{2\xi}\spin{\frac
1 M}E(x,\xi)- \frac{\Delta_\mu}{2\xi}\spin{\gamma_+}(H+E)(x,\xi)+
\nonumber\\&& +\int_{-1}^{1}du\
G_\mu(u,\xi)W_{+}(x,u,\xi)+i\epsilon_{\perp \mu k}
\int_{-1}^{1}du\  \widetilde G^k (u,\xi)W_{-}(x,u,\xi)\, ,
\nonumber \\[4mm] \widetilde F^{WW}_\mu(x,\xi)&=&
\Delta_\mu\frac12 \spin{\frac{\gamma_5}{M}}\widetilde E(x,\xi)-
\frac{\Delta_\mu}{2\xi}\spin{\gamma_+\gamma_5}\widetilde H(x,\xi)+
\nonumber\\ && +\int_{-1}^{1}du\ \widetilde
G_\mu(u,\xi)W_{+}(x,u,\xi)+i \epsilon_{\perp \mu k}
\int_{-1}^{1}du\  G^k (u,\xi)W_{-}(x,u,\xi) \nonumber \ee The
following notations are used:

\be{Gapp} G^\mu(u,\xi)&=& \spin{\gamma^\mu_\perp}(H+E)(u,\xi)+
\frac{\Delta_\perp^\mu}{2\xi} \spin{\frac 1
M}\biggl[u\frac{\partial}{\partial u}+ \xi\frac{\partial}{\partial
\xi} \biggl] E(u,\xi)- \nonumber\\[4mm]&&
-\frac{\Delta_\perp^\mu}{2\xi}
\spin{\gamma_+}\biggl[u\frac{\partial}{\partial u}+
\xi\frac{\partial}{\partial \xi}\biggl] (H+E)(u,\xi) \nonumber \ee
\be{tGapp} \widetilde G^\mu (u,\xi)& =
&\spin{\gamma^\mu_\perp\gamma_5} \widetilde H(u,\xi)
+\frac12\Delta_\perp^\mu \spin{\frac{\gamma_5}{M}}
\biggl[1+u\frac{\partial}{\partial u}+\xi\frac{\partial}{\partial
\xi}\biggl] \widetilde E(u,\xi)- \nonumber\\[4mm]&&
-\frac{\Delta_\perp^\mu}{2\xi}\spin{\gamma_+\gamma_5}
\biggl[u\frac{\partial}{\partial u}+\xi\frac{\partial}{\partial
\xi}\biggl] \widetilde H(u,\xi) \nonumber \ee

\be{Wpmapp} W_{\pm}(x,u,\xi)&=& \frac12\biggl\{
\theta(x>\xi)\frac{\theta(u>x)}{u-\xi}-
\theta(x<\xi)\frac{\theta(u<x)}{u-\xi} \biggl\} \nonumber
\\[4mm]&&\mskip-10mu \pm\frac12\biggl\{
\theta(x>-\xi)\frac{\theta(u>x)}{u+\xi}-
\theta(x<-\xi)\frac{\theta(u<x)}{u+\xi} \biggl\}. \nonumber \ee
The sandwiching between nucleon Dirac spinors is denoted by
$\spin{\ldots}=\bar U(p')\ldots U(p)$ and the quarks flavour
dependence in the amplitude can be easily restored by the
substitution: \be{flavapp} F_\mu (\widetilde F_\mu) \to
\sum_{f=u,d,s, \dots}e_f^2\ F_\mu^f(\widetilde F_\mu^f) \, . \ee

\end{document}